\documentclass[doublecol]{epl2} 
\newcommand{\be}{\begin{equation}}
\newcommand{\ee}{\end{equation}}
\newcommand{\bea}{\begin{eqnarray}}
\newcommand{\eea}{\end{eqnarray}}
\newcommand{\ph}{\tilde{p}}

\newcommand{\pt}{\tilde{p}}
\newcommand{\ptil}{\tilde p}
\newcommand{\avg}[1]{\langle #1 \rangle}

\title{Thermodynamic behaviour of two-dimensional vesicles revisited}

\author{Mithun K. Mitra\inst{1,2} 
\and Gautam I. Menon\inst{2} 
\and R. Rajesh\inst{2}}

\shortauthor{Mithun K. Mitra \etal}
\institute{            
 \inst{1} Polymer Science and Engineering, 
University of Massachusetts 
Amherst MA 01003,  USA\\        
  \inst{2} The Institute of Mathematical Sciences, 
C.I.T. Campus, Taramani, Chennai 600113, INDIA\\

}

\abstract{We study  pressurised self-avoiding
ring polymers in two dimensions using Monte Carlo 
simulations, scaling arguments and Flory-type theories, through models 
which generalise the model of Leibler, Singh and 
Fisher [Phys. Rev. Lett. {\bf 59}, 1989 (1987)].
We  demonstrate the existence of a thermodynamic phase
transition at a non-zero scaled pressure $\ptil$, where 
$\ptil = Np/4\pi$, with the number of monomers
$N \rightarrow \infty$ and the pressure $p \rightarrow
0$, keeping $\ptil$ constant, in a class of such models. This transition is driven by bond energetics
and can be either
continuous or discontinuous. It can be interpreted as a shape
transition in which the ring polymer takes
the shape, above the critical pressure, of a regular N-gon whose 
sides scale smoothly with pressure, while staying unfaceted below
this critical pressure.
Away from these limits,  we argue that the
transition is replaced by a sharp crossover. 
The area,  however, scales with $N^2$ for all  positive $p$ in all such
models, consistent with  earlier scaling theories. }

\begin{document}

\maketitle
Fluid  vesicles in three dimensions exhibit  diverse  shapes. 
This diversity originates in the nontrivial ground state configurations 
exhibited by a simple curvature Hamiltonian for a closed surface as
the enclosed volume and surface area are varied \cite{peliti}. Not much
is known about  the behaviour of such a model system at 
non-zero temperature. However, equivalent models in 
one lower dimension should provide useful insights into behaviour in the 
experimentally relevant three-dimensional
case \cite{leibler87,peliti}.  The simplest 
such model is that of  a self-avoiding  polymer ring whose enclosed area 
couples  to a pressure difference 
term \cite{fisher66,leibler87,fisher89,camacho90}.  
 
Leibler, Singh and 
Fisher (LSF) \cite{leibler87} studied such a pressurised polymer model, defining the ring
in terms of  tethered discs. The energetics of the model
derived from a pressure term conjugated to the enclosed area.
Monte Carlo simulations  and a scaling analysis about zero pressure were
used to demonstrate that the 
radius of gyration $R_G$ and the averaged area $A$ obeyed the scaling forms
 $R_G^2 \sim N^{2\nu} X({\bar p} N^{2\nu})$ and $A \sim
N^{2\nu} Y({\bar p} N^{2\nu})$, where $\nu = 3/4$ 
exactly in $d=2$ and $X(x)$ and $Y(x)$ are scaling functions of a single scaling
variable $x$.  The dimensionless measure of the pressure difference is 
${\bar p} = p a^2/k_B T$, where $p$ is the pressure difference, $a$ is the
diameter of the disc and $T$ is the temperature, 
with the  appropriate scaling variable defined as
$x = {\bar p} N^{3/2}$.  In a scaling regime around $x
\simeq 0$, data collapse for a range of system sizes was
observed by LSF, with a $x=0$ transition separating branched
polymer behaviour, for $p < 0$, from a $p > 0$ regime
governed by the statistics of self-avoiding walks.

The LSF scaling description  applies for finite
$pN^{3/2}$.   At large (positive) pressures, it is reasonable to
expect that  $A
\sim N^2$, with the expanded shape approaching a circle.  
However, extending LSF-type scaling arguments to the large pressure case
yields $A \sim N^3$ for large pressures, assuming only that  
$pN \ll 1$\cite{maggs90}. 
Any expanded regime with $A \sim N^2$ is thus
{\em inaccessible} through  an expansion about the 
LSF fixed point at ${\bar p} = 0$.   Related scaling arguments, based on a blob 
picture, have recently been presented for the case  in which
the scaled pressure $\ptil =  Np/4\pi$ is finite and non-zero
for large $N$\cite{maggs90,haleva08}. These 
suggest  that self-avoiding pressurised chains should
swell smoothly as $\ptil$ is increased, ruling out a phase 
transition at nonzero $\ptil$.  

However, there are models for pressurised  polymer rings which {\em do} exhibit a  
continuous pressurisation transition at  non-zero  $\ph$, (although these models 
allow for self-intersections)\cite{rudnick91,gaspari93}.  
Recent work on ring polymers with a pressure term conjugated to the
algebraic (signed) area, obtains a  continuous transition,
between a collapsed phase  with $\langle A \rangle \sim N$ and
an expanded phase with $\langle A \rangle \sim N^2$ \cite{haleva06,mitra08}. 
The critical  pressure, in a Flory-like theory, arises from the competition between an 
entropic $R^2/N$ term,  which favours more
compact configurations, and a pressure
term $-\ph R^2/N$, which favours expanded configurations.
Such terms should also be present in theories 
for the self-avoiding case,  supplemented
by additional terms accounting for self-avoidance. 
Whether it is the true or the algebraic area which enters the
Hamiltonian should also be irrelevant in the large pressure limit \cite{haleva06}. 
Thus,  it is natural to ask whether the inclusion of self-avoidance 
suffices to destroy the non-zero $\ph$ transition altogether or whether some
trace of the singular behaviour at $\ph \sim 1$ in the self-intersecting case 
survives.  If it does, this would indicate a
hitherto unanticipated feature of such pressurised polymer ring 
models, with possible applications to the three-dimensional case.

\begin{figure}
\begin{center}
\includegraphics[width=0.8\columnwidth]{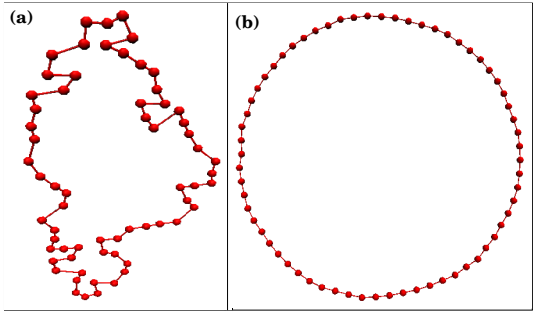}
\caption{Configurations of a self-avoiding pressurised polymer, with finite
beads connected by quadratic springs of maximum extension $R_0$ (Model C).
These figures illustrate sample configurations on either side of the transition; both for (a) $\ptil < 1$ as well as for (b) $\ptil > 1$, where 
$\ptil$ is  the critical pressure. }
\label{bondESAR} 
\end{center}
\end{figure}

This Letter demonstrates the 
existence of an unusual and somewhat subtle phase transition as $\ptil$ is increased across
a critical value,
in some models for pressurised, self-avoiding  polymer rings. This transition can be 
either discontinuous or continuous, and
separates a weakly inflated phase from a strongly inflated phase, in a sense
we make precise below. The average area  $\avg{A}$ scales as
$N^2 $ on  both sides of the transition, in accord with earlier scaling arguments.
Crucially,  the {\em nature} of fluctuations changes across  the transition 
point, with the ring expanding without feeling the effect of the maximum extension $R_0$ below the
transition, while being constrained by $R_0$ at and above the critical pressure. 
Sample configurations below and above the transition are shown in
Fig.~\ref{bondESAR}.
Interestingly,  for models of the LSF type, such transitions appear to be 
absent in the  thermodynamic limit, although their effects  may be manifest 
in sharp crossovers.

We first summarise prior scaling arguments for this problem\cite{maggs90,haleva08}. These use
the blob picture, describing the 
polymer ring as  a one-dimensional object fluctuating in two dimensions. In response to a
relevant perturbation, the ring can be naturally decomposed  into $N/g$ blobs of g monomers 
each. A blob  stores an average tensile energy equal to
$k_B T = 1$, i.e. $\gamma \xi \sim 1$, where $\gamma$ is the induced surface 
tension due to the perturbation and $\xi$ is 
the blob size. At length scales smaller than the blob size
$\xi$, the ring  is unaffected by the perturbation,   
thus obeying $\xi \sim g^{\nu}$, where $\nu$ is the swelling exponent characterising
the unperturbed manifold statistics. This implies that the surface tension
scales as $\gamma \sim g^{-\nu}$. At length scales larger than $\xi$, the
perturbation stretches the manifold;  the total projected length of the 
chain is the product of the number of blobs  and the projected length of each 
blob:  $R \sim \left( \frac{N}{g} \right) \xi \sim N \gamma^{1/\nu - 1}$.
For pressurised rings, the surface tension and  the pressure difference $p$ 
are related via Laplace's
Law, $\gamma/R \sim p$. The total projected length then  scales as
$R \sim N^{\frac{\nu}{2\nu -1}} p^{\frac{1-\nu}{2\nu -1}}$,
while the average area scales as 
\be
\avg{A} \sim R^2 \sim N^2 (pN)^{\frac{2-2\nu}{2\nu-1}} ,
\ee
with the surface tension given by $\gamma \sim p R \sim (pN)^{\frac{\nu}{2\nu-1}}$.

The above scaling argument, while justified for $pN \ll 1$, must be 
carefully examined for   $pN \sim 1$, 
{\it i.e.} when the blob size and  the microscopic cutoff are comparable.
For the blob picture to be valid,  the number of blobs should be  larger 
than unity as well as  less than $N$; these conditions impose 
$N^{-2 \nu} < p < N^{-1} $, or equivalently $pN < 1$.

The fact that  such an analysis predicts a leading $ \avg{A} \sim N^2$ behaviour,  as is physically
reasonable in the limit of large pressures, suggests that it is reasonable
to assume that the ring should 
swell smoothly at {\em all} pressures, with
the average area and surface tension  described by the general scaling forms 
$\avg{A} = N^2 f_1 (pN) , \gamma = f_2 (pN)$\cite{gomper92,haleva08}. This argument  also accounts for
the possibility of a phase transition in the self-intersecting case,  since the exponent of the argument in the 
scaling functions above
diverges at $\nu = 1/2$. Numerical simulations of a specific generalisation of the LSF model
support the conjecture that no phase transitions occur upon varying $\ptil$  in this model and, 
by extension, in the general case~\cite{haleva08}.

\begin{figure*}[t]
\begin{center}
\includegraphics[width=1.95\columnwidth]{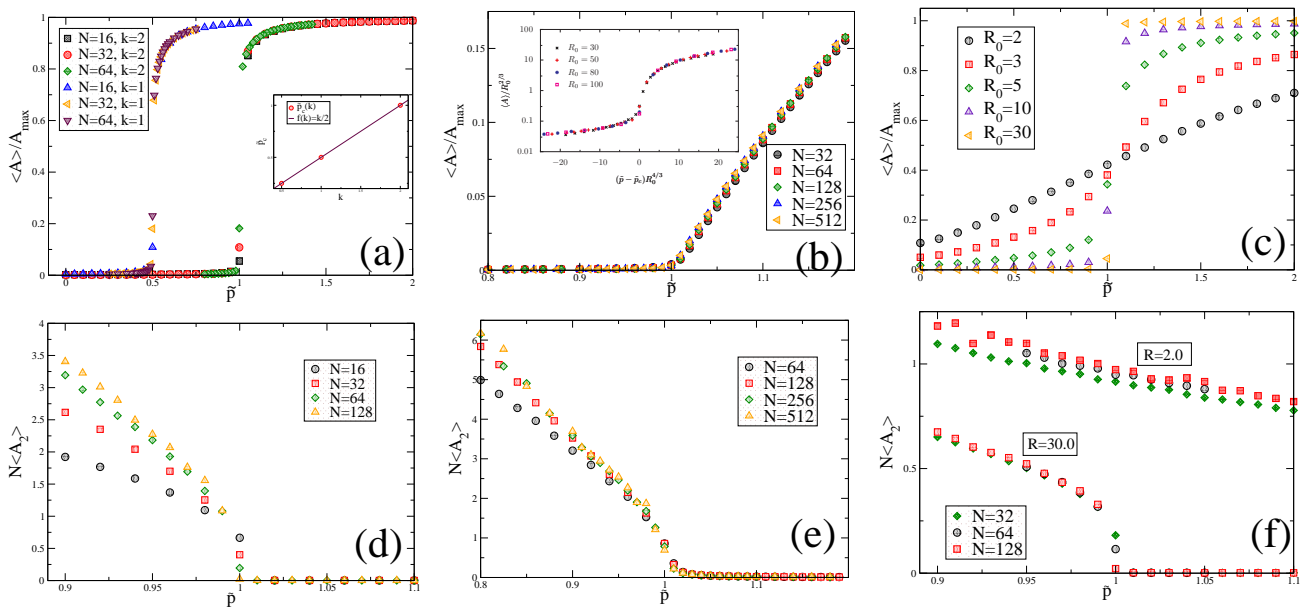}
\caption{Order parameter $\frac{A}{A_{max}}$ for (a) [Model A], (b) [Model B] and (c) [Model C]
   	and the scaled asphericity $NA_2$ for (d)  [Model A], (e) [Model B] and (f) [Model C],
illustrating their variation with the scaled pressure $\ph$. Model A, corresponding to (a) and (d), 
shows a sharp discontinuous jump in the order parameter, mirrored to
a lesser extent in the asphericity. The inset to (a) shows the proportionality $\ph_c = k/2$ for Model A.
For Model B, corresponding to (b) and (e), both quantities vary continuously
across the transition. The inset in (b) shows collapse of the data for
different $R_0$, when $\ph - \ph_c$ and $\langle A \rangle$ are scaled as in
Eq.~\ref{eq:area_col}.
For Model C, as shown
in (c) and (f), the transition can be sharp or non-existent depending on the ratio $\delta = a/R_0$. This 
behaviour can be seen both in the order parameter as well as the asphericity.}
\label{fig:area_asph}
\end{center}
\end{figure*}

Consider, however, Figs.~\ref{fig:area_asph}(a) and (b). Both represent the variation of an
appropriately normalised area
as a function of a scaled pressure variable,
in two different models of pressurised, self-avoiding polymer rings. As is apparent,
the   average area varies discontinuously in one model 
(Fig.~\ref{fig:area_asph}(a)),
whereas the other (Fig.~\ref{fig:area_asph}(b)) appears to show a continuous transition. 

Such  behaviour is precluded by the scaling arguments discussed above, but it is
easy to reason that the discrepancy originates in the extension of the blob picture to regimes 
where it is inapplicable.  When $\ph \sim 1$, the energetics of the linkages
connecting the monomers is probed\cite{maggs90}. Equivalently,  the blob size reduces to the size of the
monomer in this limit. Thus $\nu = 1$ since the ring is no longer fractal at
or below this scale, and the exponent $(2-2\nu)/(2\nu-1)$  
in the expansion $\avg{A}  \sim N^2 \left 
[ (pN)^{\frac{2-2\nu}{2\nu-1}} \right ]$ vanishes. 
Sub-leading corrections  from perimeter contributions  dependent on the detailed potential
between particles constituting the ring, must then be accounted for.

\section{Models}
We define and study
three models of pressurised self-avoiding ring polymers.
The first, model A,  consists of point particles connected by quadratic springs
with a maximum extension of $R_0$. The springs cannot intersect, thus enforcing self-avoidance. The potential between the
particles takes the form
\bea
V(r) &=& \frac{1}{2} k r^2, \quad r < R_0, \nonumber  \\
&=& \infty, \quad r \ge R_0 .
\eea
The second, model B,  again assumes point particles connected by 
springs with a maximum extension $R_0$. However, the spring potential in this 
case is chosen to be of the Finitely Extensible Nonlinear Elastic (FENE) 
\cite{warner72} type and is given by
\bea
V(r) &=& -R_0^2 \ln \left( 1 - \frac{r^2}{R_0^2} \right), \quad r < R_0, \nonumber  \\
&=& \infty, \quad r \ge R_0 
\eea
This form of the potential allows us to investigate the effect of a smooth divergence to
infinity for $r$ approaching $R_0$, as opposed to the 
jump imposed  in model A.

Our model C consists of monomers 
of finite diameter $a$ connected by springs with a maximum length
$R_0$. Self-avoidance implies that that no bead overlaps with another bead
and that no bonds intersect each other. The spring potential is a 
quadratic potential with a cutoff at the maximum bond length $R_0$, as in
model A.  The relevant dimensionless parameter in this model is the ratio of the bead size to the
maximum bond-length,  $0 \le \delta \equiv \frac{a}{R_0} < 1$.
In the limit $\delta = 0$, we recover  model A, whereas for
$\delta = 1/1.8$, we obtain the analog of the LSF model, albeit with springs
instead of tethers. The spring constant is chosen to be $k=2$;
results for other values of $k$ are obtained 
by rescaling  the pressure axis as $\ph \rightarrow 2 \ph/k$ (see below).

\section{Simulation Details and Measured Quantities}

\begin{figure*}[t]
\includegraphics[width=2\columnwidth]{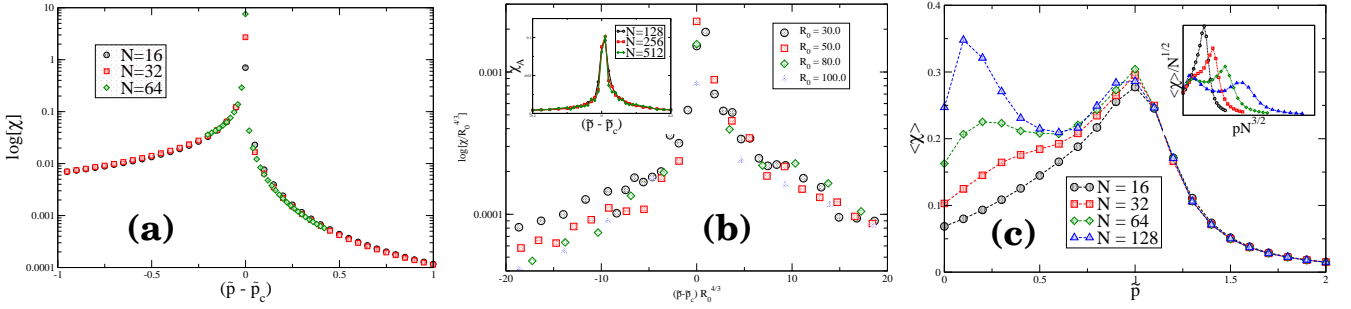}
\caption{Susceptibility plots for (a) model A, (b) model B and (c) model C. The first-order
character of the area susceptibility of model A is shown in (a), with a delta-function 
singularity at the transition.  In contrast, model B displays a power-law divergence of the susceptibility
for different values of the bond length ($R_0$), shown in (b). The inset to (b) shows the 
susceptibility of model B for different values of the system size, which all collapse onto
a single curve. The susceptibility for model C is shown for (c) for an intermediate value of
the bond length ($R_0 = 3$). The plot shows two clear peaks, 
the second peak scaling with $pN$,
and the first with $pN^{3/2}$ as shown in the inset to (c).}
\label{fig:suscept} 
\end{figure*}

We have performed  Monte-Carlo simulations
of  models A, B and C , combining both local (single bead)
and non-local (multiple bead) moves \cite{koniaris94}. 
The basic Monte Carlo move consists of choosing
a bead at random and giving it a random displacement.
Self-avoidance is ensured by  checking that the random
displacement does not lead to bond intersections or bead overlaps, if applicable \cite{madras90}.
The system is  evolved typically for $\sim 4 \times 10^6$
Monte Carlo steps, with the relevant averages  computed after 
discarding the first $10^6$ steps for equilibration. We have also
performed multi-particle moves involving the reflection of a group of
$n$ connected particles across the line joining the particles at the end.
A variety of initial configurations are used to initiate the simulations.
Defining bead positions on an underlying fine grid speeds up the simulations
considerably.

 We define the 
order parameter as the ratio of the average area to the maximum 
allowed area, $\Phi = A/A_{max}$ with
$A_{max} = \frac{1}{4} N R_0^2 \cot (\frac{\pi}{N}) $, as appropriate
to an $N$-gon of fixed side length $R_0$.
In the thermodynamic limit, defined as the double limit 
$N \rightarrow \infty$ and $R_0 \rightarrow \infty$, 
this order parameter should be zero for  $\pt < \pt_c$
and nonzero above it. 

The nature of the transition can be characterised through an 
appropriately defined susceptibility. We define the area susceptibility\cite{haleva06}
\be
\chi = \frac{1}{\avg{A}} \frac{\partial{\avg{A}}}{\partial{\pt}}.
\ee
For a continuous transition, both the order parameter and
susceptibility collapse around the critical point can be
described by appropriately defined scaling functions
\bea
\Phi &=& R_0^{\theta_1} h_1[(\pt - \pt_c) R_0^{\theta_2}], \label{eq:area_col}\\
\chi &=& R_0^{\zeta_1} h_2[(\pt - \pt_c) R_0^{\zeta_2}] .
\label{eq:sus_col}
\eea

The instantaneous shape of the ring is characterised by the gyration tensor, defined as
$
T_{\alpha,\beta} = \frac{1}{N} \sum_{i=1}^N (X_{i,\alpha} - X_{CG,\alpha})(X_{i,\beta} - X_{CG,\beta}),
$
with $X_{CG,\alpha}$ denoting the $\alpha^{th}$ component of the centre of 
mass and $X_{i,\alpha}$ denoting the $\alpha^{th}$ component of the $i^{th}$
particle.
If the eigenvalues of this  tensor are  $\lambda_1$ and
$\lambda_2$, the radius of gyration $ R^2_G = \lambda_1 + \lambda_2$ and the  asphericity $A_2$ is
\be
A_2 = \left( \frac{\lambda_1 - \lambda_2}{\lambda_1 + \lambda_2}\right)^2.
\ee
This asphericity is $0$ for a perfectly spherical shape and $1$ for a rod-like
shape.

\section{Results}

We first discuss  results for model A.
The order parameter shows a first order jump for the system sizes shown, in which
the number of monomers $N$ ranging from 16 to 64, as well as for
two different values of the spring constant $k$. At smaller $k$, the
transition $\ph$ shifts to smaller values; as shown in the inset,
the critical scaled pressure equals $k/2$. There are no
significant finite size effects in the data, with the data for $N=16$
overlapping the data for $N=64$; this is also the case  for much larger
systems (not shown). Fig.~\ref{fig:area_asph}(d) shows the asphericity across the 
transition; while $A_2$ vanishes asymptotically as $N$ is increased both above and
below the transition indicating that the asymptotic shape is a circle in both cases, there
is a significant difference in the $N$-dependence of $A_2$ on both sides of the transition. 
The susceptibility (Fig.~\ref{fig:suscept}(a)) displays
a classic first-order delta-function peak at the transition (note the logarithmic
scale on the y-axis) with the peak height increasing with $N$.

The dependence of the critical pressure on $k$ can be computed in the
following way, relying on the assumption that fluctuations can be ignored 
for $\ptil > \ptil_c$: The pressurisation energy is
$F_P = -4 \pi^2 \ptil \frac{R^2}{N}$. Adding the spring energy 
$F_S = \frac{N}{2} k (\frac{2 \pi R}{N})^2 = 2 \pi^2  k \frac{R^2}{N}$ to this
gives $F_P + F_S = \left [ -2\ptil  + k \right ] \frac{R^2}{N}$ and  thus a 
first-order transition at  $\ptil_c = k/2$, as seen in the data. The 
self-avoidance term will, of course, prevent collapse 
for $\ptil <  \ptil_c$; a Flory theory discussed below indicates that $\avg{A} \sim N^2$
in this phase. 

\begin{figure}[t]
\begin{center}
\includegraphics[width=0.8\columnwidth]{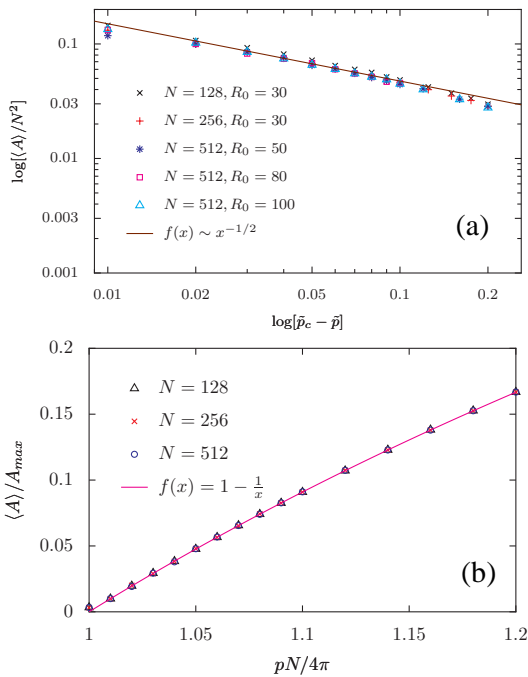}
\caption{(a)The divergence of area with pressure near the critical point for a
range of system sizes and maximum bond lengths for model B; (b) The large 
pressure behaviour of the average area for model B. The data is shown for
three system sizes N = 128, 256, 512 with $R_0$ = 30.0 .}
\label{fig:p_behav} 
\end{center}
\end{figure}

In contrast, order parameter results for model B (Fig.~\ref{fig:area_asph}(b)), indicate
a smooth increase from zero at the critical pressure value. The order
parameter curves show no  dependence on $N$ but depends on
$R_0$ above the transition. The asphericity
(Fig.~\ref{fig:area_asph}(d)) varies strongly 
with $N$ below the transition, while the $N$-dependence essentially 
collapses above the transition. The susceptibility plots for model B
are shown in Fig.~\ref{fig:suscept}(b). Data for different values of $R_0$ can be
collapsed using Eq.~\ref{eq:sus_col},
where the exponents $\zeta_1 \approx 1.33$ and $\zeta_2 \approx 1.33$
as shown in the inset to Fig.~\ref{fig:area_asph}(b).
The order parameter variation can also be collapsed using the
form given by Eq.~\ref{eq:area_col} with exponents given by
$\theta_1 \approx 0.67$ and $ \theta_2 \approx 1.33$. (The errors on these
exponents are large, however, and conservative estimates of 
error bars are around $\pm 0.10$ on each of the calculated exponents.) 
For $\ph < \ph_c$, we
have $\frac{\avg{A}}{N^2} \sim (\pt_c - \pt)^{-\theta_1/\theta_2}$, as
illustrated in Fig.~\ref{fig:p_behav}(a) . Above the transition, the area
order parameter can be calculated analytically
(as shown in the next section) yielding, for $\ph > \ph_c$,
$\frac{\avg{A}}{N^2} \sim (1 - 1/\pt)R_0^2$,  as illustrated in 
Fig.~\ref{fig:p_behav}(b).

These results for self-avoiding rings with point particles then imply 
the following:   If there is a finite maximum extension $R_0$, then there is a 
transition such that $\lim_{R_0 \rightarrow \infty, N \rightarrow \infty}
\frac{A}{A_{max}}$ is zero below the transition and non-zero above it.
This transition separates a {\em weakly} expanded phase, in which the 
area scales as $N^2$ but is otherwise insensitive to the value of
$R_0$, from a {\em strongly} expanded phase, in which the area scales as $(NR_0)^2$.
Depending on the form of the potential, the transition between
these phases can either be discontinuous
or continuous.  The transition can also be characterised as a shape 
transition: below $\ptil$,  the asphericity $A_2$ is  $R_0$-independent for 
finite $N$ while it decreases to zero with increasing $R_0$ above it. 

To address the question of transitions in models of the  LSF type,
we now present our results for model C, where  the monomers are now 
beads of a finite diameter $a$ connected by springs with a maximum allowed extension $R_0$.  Order parameter plots 
for this model are shown in Fig.~\ref{fig:area_asph}(c). For the smallest $R_0$, the order
parameter varies smoothly through the transition. As $R_0$ is
increased, however, the variation becomes sharper, with 
what appears to be discontinuous behaviour of the order parameter
at large $R_0$. This behaviour is also apparent in 
the asphericity, shown for two values of $R_0$ in Fig.~\ref{fig:area_asph}(f). For 
$R_0=2$ ($\delta = 1/2$), there is no transition, while for $R_0 = 30$
($\delta = 1/30$), a sharp jump in the asphericity
across the transition is  seen. The susceptibility plots for this
model (Fig.~\ref{fig:suscept}(c))
show two distinct peaks, the first about $p=0$ scaling as $pN^{3/2}$, and the
second about $\ph = 1$, scaling as $pN$.

Thus either, (a)  there is a critical value of $\delta$ 
below which there is no transition as a function of $\ptil$ and 
above which there is a discontinuous transition or, (b) there is
no transition except in the limit of $\delta 
\rightarrow 0$ (model B), where a transition demonstrably
exists. We conjecture that the second scenario is applicable, for the following
reason: As $\ptil  \rightarrow \ptil_c$, blob sizes becomes comparable to 
the microscopic cutoff.  However,  for  finite  beads, the monomer dimension bounds
the blob size from below since fluctuations below this size cannot be resolved,
giving the perimeter a finite width even in the limit of very large pressures.
The smoothness assumption implicit in the
blob-based scaling arguments should thus continue to hold, with both 
 area and the surface tension varying non-singularly as
$p$ is increased. For models with  non-vanishing $\delta$, 
we therefore conjecture that there can only be a crossover, though possibly
a sharp one,  from the weakly to the
strongly expanded phase. This is consistent with the observed absence of the 
transition at large pressures in LSF-like models~\cite{haleva08}.

Formally,  a vesicle perimeter of mean length $L \sim 2 \pi R$ undergoes
transverse thermal wandering of a magnitude
$u_{rms} \sim  \left (2\pi k_BT/p \right )^{1/2}$ \cite{maggs90}.
Note that $u_{rms}$ is  independent
of $N$, validating a central assumption of the Flory theory presented later.
Given $\gamma = pR $, the  ratio of the intrinsic width $\xi$ (blob size) to $u_{rms}$  is
$\frac{\xi}{u_{rms}} \sim (pN)^{\frac{\nu}{2\nu - 1}} (\frac{p}{2\pi k_BT} )^{1/2}$.
Inserting  $\nu=3/4$, 
we obtain $\frac{\xi}{u_{rms}} \sim (\ptil)^{3/2} (p/2\pi k_BT )^{1/2}$.
With $p \rightarrow 0, N \rightarrow \infty$, with $\ptil = Np = {\rm const}$,
$\frac{u_{rms}}{\xi} \rightarrow \infty$; this conclusion holds true even if we assume
$\nu \rightarrow 1$. Thus, any scale 
which  bounds  the blob size $\xi$ from below, such as a bead dimension, should
provide a smooth variation of  $\avg{A}$ at all pressures.

\section{Scaling Arguments}

A Landau formulation justifying the existence of a transition can be framed in
terms of the 
competition between a surface free energy arising out of an effective
tension term $\frac{N}{2} \sigma(\frac{2 \pi R}{N})$
and the pressurisation term $-\pt\frac{R^2}{N}$. The expansion of the
surface tension term yields terms of the form$ \frac{R^2}{N} (a_0 + a_2 \frac{R^2}{N^2} 
+ a_4 \frac{R^4}{N^4} \ldots ) $, where the lowest order contribution $(a_0 > 0)$ is present even
in the tethered case,  from entropic elasticity. Combining terms of the
same order yields a Landau free energy of the form $F_L \sim \alpha_0\frac{R^2}{N}  + 
\alpha_1  \frac{R^4}{N^3}\ldots $, with the $\alpha_0$ term changing sign at a
critical value of the pressure. Such a Landau theory predicts an 
(area) order parameter exponent of 1 for the continuous case (model B), while comparing free
energies on both sides of the transitions yields first order behaviour for model A, as
seen. The nature of the  $\pt < \pt_c$  phase  requires consideration of the effects of 
self-avoidance.
 
A simple Flory-like  theory, accounting for contributions
from self-avoidance, pressurisation, entropy  and bond 
stretching,  provides a consistent explanation of the behaviour at
all pressures. The stretching free
energy is a function of  the maximum bond length $R_0$ and the
form of the potential between neighbouring monomers. For model B,
(and, in fact, fairly generically)
it is of the form $F_{stretching} \sim \frac{R^4}{R_0^2 N^3}$\cite{haleva08}.
Apart from a self-avoidance term $F_{SA}$, we may take over 
terms from the study of the pressurised self-intersecting ring,
to get  a Flory free energy
$F \simeq F_{SA} + \frac{R^2}{N} (1 -\pt) + \frac{R^4}{R_0^2 N^3}$.
To estimate $F_{SA}$, we assume that since the model is always in
an expanded phase with $\avg{A} \sim  N^2$, $F_{SA}$ is well represented by
small fixed-width fluctuations  about the N-gon shape. The contribution to the 
free energy from self-avoidance is then approximated as $F_{SA} = N^2/R$ (as
opposed to $N^2/R^2$ for a two dimensional self avoiding ring).
Minimising this  free energy yields the following predictions,
\be
\avg{A} \sim \avg{R^2} \sim \left\{ \begin{array}{ll}
	\frac{N^{2}}{(1-\pt)^{2/3}}, & \hspace{0.5cm}\mbox{$\pt < \pt_c$} \\
	N^{2} R_0^{4/5}, & \hspace{0.5cm}\mbox{$\pt = \pt_c$} \\
	N^{2} R_0^{2} (\pt-1), & \hspace{0.5cm}\mbox{$\pt > \pt_c$} 
	\end{array}
	\right.
\ee
Note that $\avg{A}$ always scales as $N^2$ and is independent of
$R_0$ below the transition.  Above the transition,  $\avg{A}  \sim  (\pt-1)$, 
increasing linearly just above the transition, consistent with Fig. 2(b) and Landau-theory
predictions. Thus, such a Flory-type scaling theory 
predicts the existence of a transition at a critical value of the scaled
pressure. The exponents for the area scaling predicted using this Flory 
theory are $\theta_1 \approx 4/5$ and $ \theta_2 = 6/5$, at the outside limit of
the error bars of the  values we obtain numerically for Model B 
(0.67 and 1.33).  However, errors on these exponents in the simulations are 
substantial and  the predictions of our Flory-type theory may possibly be correct.

The behaviour of the area above the transition can be computed analytically. 
In this regime, the shape of the polygon 
ring is simply a regular polygon and the free
energy is  simply the sum of the pressurisation energy and
the spring energy. For a regular
polygon of side $a$, the area is given by $A = \frac{1}{4} N a^2 \cot (\frac{\pi}{N}) $,
while $A_{max} = \frac{1}{4} N R_0^2 \cot (\frac{\pi}{N}) $,
where $R_0$ is the maximum allowed bond length.
For Model B,  the free energy can be written as,
$F = -pA - N R_0^2 \ln \big( 1 - \frac{A}{A_{max}} \big)$ .
The average area is obtained from $\partial F/{\partial A} = 0$, giving
$p = \frac{4 \tan(\pi/N)}{1 - \frac{\avg{A}}{A_{max}}} $.
Defining  $\tilde{p} = \frac{1}{4} p \cot (\frac{\pi}{N})$ ,
we obtain
$\pt = \frac{1}{1 - \frac{\avg{A}}{A_{max}}} $.
Note that this definition for $\pt$ is in fact the correct one, reducing to
the usual $\pt = \frac{pN}{4 \pi}$ for large system sizes. Thus 
\be
\frac{\avg{A}}{A_{max}} = 1 - \frac{1}{\pt} .
\label{eq:areaasymp_c}
\ee
This relation predicts the behaviour of
the area very accurately for large pressures and is consistent with
numerical data extending almost up to the critical pressure for
large $R_0$, where it coincides with the prediction from Flory theory:
$\avg{A} \sim N^{2} R_0^{2} (\pt-1)$ to lowest order. 
This is shown in Fig.~\ref{fig:p_behav}(b).

\section{Conclusions}
We have studied the thermodynamics of  several  models which generalise the  original model
of LSF. We demonstrate the existence of a phase transition in an appropriately scaled,
pressure-like quantity.  This  transition is best thought of as a shape transition, in which
the polymer conformation and area scaling are independent of a  parameter $R_0$
which sets the maximum bond size
below the critical pressure, while  depending strongly upon it above. 
The polymer ring is un-faceted below the transition and 
faceted in the form of an N-gon just above, swelling smoothly thereafter while retaining its faceted shape 
as the pressure is increased further above $\ph_c$. Such transitions can be discontinuous or continuous in the
limit that the ratio of the bead size to the characteristic scale of bead separation vanishes.
In the general limit, where this ratio is non-zero, we conjecture that no transition exists,
a conclusion which should apply  to models of the LSF type.

The possibility of a genuine phase transition in an appropriately scaled pressure variable
in models for self-avoiding pressurised rings has not been raised before, although the
presence of such a transition in the self-intersecting case has been settled
decisively\cite{haleva06,mitra08}.  Simple  Flory-type theories appear to give useful insights into the 
transition. A similar study for the  three-dimensional case 
may yield important insights and would appear to be very worthwhile.

\acknowledgments
We thank D. Dhar for useful discussions.

\end{document}